\documentclass[journal]{IEEEtran}
\renewcommand{\baselinestretch}{0.97}
\usepackage{pdfpages}
\usepackage{amsmath, graphics, amssymb, epsfig, float}
\usepackage[american]{babel}
\usepackage{amsfonts, graphicx, color}
\usepackage{cite, array, subfigure, multirow, hhline}
\usepackage[noend]{algpseudocode}

\def\BibTeX{{\rm B\kern-.05em{\sc i\kern-.025em b}\kern-.08em
    T\kern-.1667em\lower.7ex\hbox{E}\kern-.125emX}}

\def\FigureWidth{3.2in}
\def\FigureWidthee{3.4in}

\newtheorem{algorithm}{Algorithm}

\newcommand{\be}{\begin{equation}}

\newcommand{\ee}{\end{equation}}
\newcommand{\bea}{\begin{eqnarray}}
\newcommand{\eea}{\end{eqnarray}}
\newcommand{\bdp}{\begin{displaymath}}
\newcommand{\edp}{\end{displaymath}}
\newcommand\NoDo{\renewcommand\algorithmicdo{}}

\floatstyle{ruled}
\newfloat{algorithm}{tbp}{loa}
\providecommand{\algorithmname}{Algorithm}
\floatname{algorithm}{\protect\algorithmname}

\makeatletter
\def\hlinewd#1{\noalign{\ifnum0=`}\fi\hrule \@height #1 \futurelet \reserved@a\@xhline}

\begin{document}
\title{\LARGE{Multi-Agent Deep Reinforcement Learning for Distributed Resource Management in Wirelessly Powered Communication Networks}}
\author{\IEEEauthorblockN{Sangwon Hwang, Hanjin Kim, Hoon Lee, and Inkyu Lee,~\IEEEmembership{Fellow,~IEEE}}
\vspace{-8mm}}
\author{\IEEEauthorblockN{Sangwon Hwang, Hanjin Kim, Hoon Lee, and Inkyu Lee,~\IEEEmembership{Fellow,~IEEE}}
\vspace{-8mm}
\thanks{Copyright (c) 2015 IEEE. Personal use of this material is permitted. However, permission to use this material for any other purposes must be obtained from the IEEE by sending a request to pubs-permissions@ieee.org.

This work was supported in part by the National Research Foundation of Korea (NRF) Grant funded by the Korea Government (MSIT) under Grant 2017R1A2B3012316 and Grant 2019R1F1A1060648.

S. Hwang and I. Lee are with the School of Electrical Engineering, Korea University, Seoul 02841, Korea (e-mail: \{tkddnjs3510,inkyu\}@korea.ac.kr).

H. Kim is with Samsung Research, Samsung Electronics Co., Ltd., Seoul, Korea (e-mail: hanjin86.kim@samsung.com).

H. Lee is with the Department of Information and Communications Engineering, Pukyong National University, Busan 48513, Korea (e-mail: hlee@pknu.ac.kr).}}
\maketitle

\begin{abstract}
This paper studies multi-agent deep reinforcement learning (MADRL) based resource allocation methods for multi-cell wireless powered communication networks (WPCNs) where multiple hybrid access points (H-APs) wirelessly charge energy-limited users to collect data from them. We design a distributed reinforcement learning strategy where H-APs individually determine time and power allocation variables. Unlike traditional centralized optimization algorithms which require global information collected at a central unit, the proposed MADRL technique models an H-AP as an agent producing its action based only on its own locally observable states. Numerical results verify that the proposed approach can achieve comparable performance of the centralized algorithms.
\end{abstract}

\vspace{-1mm}
\section{Introduction}

Recently, radio frequency (RF) based energy harvesting (EH) techniques have attracted a significant attention owing to its capability for charging devices remotely \cite{RZhang:13, Bu:20, Bu:19, Chen:19}. Wireless powered communication networks (WPCNs) \cite{Bi:16,Niyato:17}, which jointly design the wireless charging and communication protocols, have been regarded as a promising solution for extending the life of the energy-constrained mobile users at the network. In the WPCNs, a hybrid access points (H-APs) broadcasts an RF signal to devices in the downlink, and the devices harvest the energy to transmit information signals in the uplink. Thus, it is important to carefully design resource management algorithms for joint optimization of information and energy transmission.

There have been intensive studies on resource management in the WPCNs for various scenarios with multiple users \cite{HLee:16,HLee:19}, and H-APs \cite{HKim_Letter:18,HKim:18,Ma:16}, which focus on charging small devices such as wireless sensors and internet of things (IoT) devices. Most of existing works, however, assume centralized computations where instantaneous full channel state information is required and the information exchange in large networks thus incurs prohibitive backhaul overhead. To avoid such difficulties, the work \cite{Ma:16} investigated distributed resource optimization approaches. However, they adopted an additional central coordinator that schedules computations of multiple H-APs. Such an assumption would not be practical in the WPCN which mainly relies on ad-hoc networking configurations, e.g., IoT systems. This motivates the development of distributed resource management policies for the WPCN without additional centralized units, which is, in general, challenging for conventional optimization methods.

This paper investigates a deep reinforcement learning (DRL) approach for the WPCN which allows distributed calculations at individual H-APs. The DRL has been recently applied to solve resource allocation problems in various wireless systems \cite{APaul_TVT:20,Wu:19,Anh:19,Huynh:16,Yaohua:19,Chaofan:19,Yasar:19,Le:19}. The work in \cite{APaul_TVT:20} mitigates the end-to-end outage probability in an EH enabled cognitive radio networks through RL based Q-routing. A multi-agent DRL (MADRL) architecture has been extended in \cite{Yasar:19} and \cite{Le:19}. In the MADRL framework, a network entity responsible for individual computations can be modeled as an agent that determines an action, i.e., a resource management policy, using locally observable states along with interactions with other agents. The work in \cite{Yasar:19} employed the MADRL method to develop distributed computation strategies of power allocation solutions in an ad-hoc setup. Also, \cite{Le:19} presented MADRL-based spectrum sharing schemes for vehicular networks, which adopted a distributed execution process of centrally trained agents. The agents are thus optimized with the aids of centralized computations, but their real-time realizations are carried out individually. Such a concept might not be feasible for practical WPCNs where no central coordinator is allowed even in the training step due to arbitrarily deployed H-APs. Therefore, for the WPCN scenarios, it is essential to develop a new MADRL framework that enables decentralized inferences both in the training and execution steps.

This paper proposes a MADRL-based distributed resource allocation strategy for maximizing the sum-rate performance in the multi-cell WPCN where the agents implemented at the H-APs are trained and executed in a distributed manner. To this end, each agent is realized by its own deep neural network (DNN) that can be individually learned without knowing global information of the system. In particular, an agent leverages locally observable knowledge which can be sensed from other cells, i.e., power of interference and energy signals. Such a local state can be successfully maintained in a fully distributed manner. To prevent egoistic resource allocation strategies, we design a local reward so that each H-AP can individually determine its networking policy with distributed coordination among other H-APs during the training.

Consequently, the proposed MADRL approach accomplishes not only the distributed training with the aid of a simple interaction among the agents, but also the distributed execution using only locally observable information. Furthermore, the proposed strategy can reduce computational complexity compared with the conventional centralized algorithm in \cite{HKim:18}. Numerical results verify that the proposed distributed resource allocation policy achieves comparable performance of conventional centralized computation systems with lower computational complexity. The contributions of this paper are summarized as follows:
\begin{itemize}
    \item The MADRL-based distributed optimization method is proposed for multi-cell WPCN systems. An RL reformulation of the sum-rate maximization is introduced by carefully designing actions, states, and rewards suitable for the WPCN.
    \item An efficient interaction mechanism among H-APs is developed to accomplish decentralized training and execution of agents.
    \item The effectiveness of the proposed approach is demonstrated in comparison of traditional centralized optimization algorithms and state-of-the-art MADRL solutions.
\end{itemize}
\vspace{-3mm}

\vspace{-1mm}
\section{System Model And Problem Formulation} \label{sec:main1}

\begin{figure}
\begin{center}
\includegraphics[width=\FigureWidth]{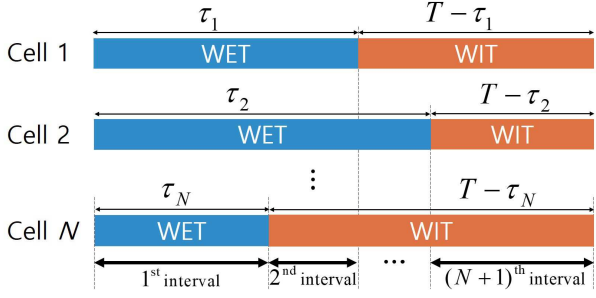}
\end{center}
\caption{Illustration of the $N$ cell WPCN operation}
\label{figure:system}
\vspace{-1mm}
\end{figure}

\begin{table}[t]
    \caption{List of Symbols}
    \label{table:Symbols}
    \centering
\begin{tabular}{|c|l|}
\hline
Symbol & Definition\\\hline
$h_{ij}^{(t)}$ & Channel gains from user $i$ to H-AP $j$\\
$g_{ij}^{(t)}$ & Channel gains from H-AP $i$ to H-AP $j$\\
$b_{ni}^{(t)}$ & Binary number indicating if H-AP $i$ receives WIT signal\\
$I_{nji}^{(t)}$ & WIT interference from cell $j$ in interval $n$\\
$D_{nji}^{(t)}$ & Cross-link WET interference from cell $j$ in interval $n$\\
$\tau_{i}^{(t)}$ & Duration of the WET operation at H-AP $i$\\
$p_{i}^{(t)}$ & Transmit power of user $i$\\
$s_{i}^{(t)}$ & State of H-AP $i$ at time slot $t$\\
$s_{i,\text{I}}^{(t)}$ & Internal state of H-AP $i$ at time slot $t$\\
$s_{ji,\text{E}}^{(t)}$ & External state of H-AP $i$ at time slot $t$ sensed from cell $j$\\
$\hat{E}_{ji}^{(t)}$ & Estimation of the harvested energy\\
$\hat{I}_{ji}^{(t)}$ & Estimation of the WIT interference\\
$\hat{D}_{ji}^{(t)}$ & Estimation of the cross-link WET interference\\
\hline
\end{tabular}
\vspace{-4mm}
\end{table}

We consider a multi-cell WPCN scenario in \cite{HKim:18} where $N$ users wish to communicate with their corresponding H-APs. The symbols which are used throughout the paper are summarized in Table \ref{table:Symbols}. The H-APs first charge the users by transmitting energy-carrying RF signals in the downlink wireless energy transfer (WET) phase, and the users harvest the energy from the received signals. Then, by exploiting the harvested energy, the users transmit their information signals to the H-APs in the uplink wireless information transmission (WIT) phase.

\begin{figure}
\begin{center}
\includegraphics[width=\FigureWidthee]{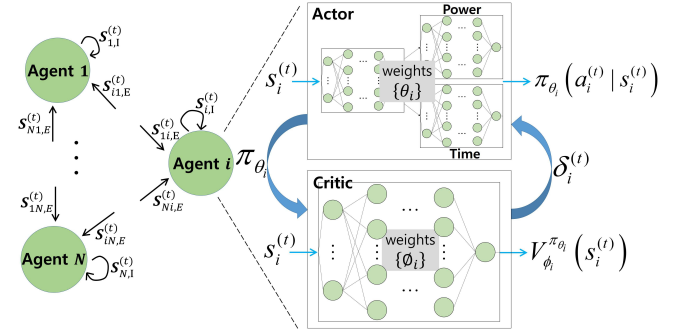}
\end{center}
\caption{Network architecture of MA-A2C}
\label{figure:network}
\vspace{-3mm}
\end{figure}

We briefly explain the operation of the multi-cell WPCN system, as shown in Fig. \ref{figure:system}. The WPCN process is carried out in the time-slotted manner where at time slot $t$ of duration $T$, H-AP $i$ first performs the WET in the downlink of duration $\tau_{i}^{(t)}$. Then, the WIT of user $i$ is conducted in the uplink during the remaining $T-\tau_{i}^{(t)}$. Then, assuming $N$ cells, as illustrated in Fig. \ref{figure:system}, the total system block is divided into $N+1$ intervals. Since the operation of the overall system is different at each interval, the rate and the harvested energy performance should be carefully characterized. To this end, an ordering $\mu_{i}$ of the H-APs is defined as $\tau_{\mu_{1}}\leq\cdots\leq\tau_{\mu_{N}}$. Then, the duration of interval $n$ ($n=1,\cdots,N+1$) becomes $\tau_{\mu_{n}}^{(t)}-\tau_{\mu_{n-1}}^{(t)}$ with $\tau_{\mu_{0}}\triangleq0$ and $\tau_{\mu_{N+1}}\triangleq T$. 
Assuming the time-slotted block fading model, at time slot $t$, the uplink channel from user $i$ to H-AP $j$  ($i,j=1,\cdots,N$) is denoted by $\tilde{h}_{ij}^{(t)}$. Based on the Jake's model \cite{Yasar:19}, the channel coefficient follows a first-order complex Gauss Markov process $\tilde{h}_{ij}^{(t)}=\rho\tilde{h}_{ij}^{(t-1)}+ \sqrt{1-\rho^{2}}e_{ij}^{(t)}$
where $\rho = J_{0}(2\pi f_{d}T)$ stands for the time correlation of the channel, $J_{0}(\cdot)$ is the zeroth-order Bessel function of the first kind, $f_{d}$ represents the maximum Doppler frequency and $e_{ij}^{(t)}\sim\mathcal{CN}(0,1)$ indicates the channel innovation process that is independent of $\tilde{h}_{ij}^{(t)}$. In a similar way, we define the channel coefficient $\tilde{g}_{ij}^{(t)}$ from H-AP $i$ to H-AP $j$ as the Gauss Markov process.

At interval $n$, H-AP $i$ with $\tau_{\mu_{n}}^{(t)}>\tau_{i}^{(t)}$ receives the WIT signals transmitted from users, whereas those with $\tau_{\mu_{n}}^{(t)}\leq\tau_{i}^{(t)}$ transfer the energy in the downlink. For convenience, let $b_{ni}^{(t)}$ be a binary number indicating the mode of H-AP $i$ at interval $n$ such that $b_{ni}^{(t)}=1$ if H-AP $i$ receives the WIT signal and $0$ otherwise. Notice that $b_{ni}^{(t)}$ is not an optimization variable and is straightforwardly determined by the time allocation variables. For $b_{ni}^{(t)}=1$, H-AP $i$ experiences WIT interference from other user $j$ with $b_{nj}^{(t)}=1$ and cross-link WET interference from H-AP $j$ with $b_{nj}^{(t)}=0$.

Let us denote $p_{i}^{(t)}$ and $P$ as the uplink transmit power of user $i$ and the downlink power at the H-APs, respectively. Also, we define $h_{ij}^{(t)}\triangleq|\tilde{h}_{ij}^{(t)}|^2$ and $g_{ij}^{(t)}\triangleq|\tilde{g}_{ij}^{(t)}|^2$ as the channel gains from user $i$ to H-AP $j$ and from H-AP $i$ to H-AP $j$, respectively. Then, the instantaneous WIT interference $I_{nji}^{(t)}$ from user $j$ to H-AP $i$ and the instantaneous cross-link WET interference $D_{nji}^{(t)}$ from H-AP $j$ to H-AP $i$ at interval $n$ are respectively expressed as
\begin{align}\label{eq:interference}
    I_{nji}^{(t)}=h_{ji}^{(t)}p_{j}^{(t)}b_{nj}^{(t)}\ \text{and}\ D_{nji}^{(t)}=\beta g_{ji}^{(t)}P(1-b_{nj}^{(t)})
\end{align}
with the attenuation factor $\beta$.\footnote{If the WET signals are shared among H-APs, the WET interference $D_{nji}^{(t)}$ is perfectly canceled as $\beta = 0$. Otherwise, we have $0<\beta<1$~\cite{HKim:18}.} Since the uplink data transmission of user $i$ at interval $n$ is carried out when $b_{ni}^{(t)}=1$, the corresponding achievable data rate $R_{ni}^{(t)}$ can be written by
\begin{align} \label{crosslinkrate}
R_{ni}^{(t)}\!=\!(\!\tau_{\mu_{n}}^{(t)}\!-\!\tau_{\mu_{n\!-\!1}}^{(t)}\!)b_{ni}^{(t)}
\log\!\!\Bigg(\!\!1\!+\!\frac{h_{ii}^{(t)}p^{(t)}_{i}}{\sigma^{2}\!+\!\sum_{j\neq i}(I_{nji}^{(t)}\!+\!\!D_{nji}^{(t)})}\!\Bigg),
\end{align}
where $\sigma^{2}$ equals the noise power. The multi-user WIT interference $I_{nji}^{(t)}$ appears in \eqref{crosslinkrate} since all  users share the same time-frequency resources. Then, the total achievable rate of user $i$ over $N+1$ intervals is denoted by $R_{i}^{(t)}=\sum_{n=1}^{N+1}R_{ni}^{(t)}$.

On the other hand, user $i$ harvests the energy of the received RF signals when $b_{ni}^{(t)}=0$. The contribution of H-AP $j$ on the harvested energy of user $i$ at interval $n$ is given as

\vspace{-4mm}
\begin{align}
    \color{black}
    E_{nji}^{(t)}=\triangle\Big(P h_{ij}^{(t)}(1-b_{ni}^{(t)})(1-b_{nj}^{(t)})\Big),\label{Ein}
\end{align}
where the function $\triangle(x)$ defines the input-output relationship of EH circuits for a given input power $x$ and $1-b_{nj}^{(t)}$ appears since H-AP $j$ can only affect the EH performance of user $i$ if it radiates the downlink WET signals. The harvested energy of an ideal EH model is given as $\triangle(x)=\eta x$
with $\eta\in(0,1]$ being the energy harvesting efficiency. Also, the non-linearity of practical EH circuits is modeled as~\cite{Skang:19}
\begin{align}\label{eq:EH_NL}
    \triangle(x)\!=\!\frac{a_{3}(1-\exp{(a_{1}x)})}{1+\exp{(-a_{1}x+a_{2})}},
\end{align}
where $a_{k}$ for $k=1,2,3$ are fitting parameters. Thus, the total harvested energy at user $i$ over all intervals is written by $E_{i}^{(t)}\triangleq\sum_{n=1}^{N+1}(\tau_{\mu_{n}}^{(t)}-\tau_{\mu_{n-1}}^{(t)})\sum_{j=1}^{N}E_{nji}^{(t)}$. Notice that the transmit EH consumption of user $i$ cannot exceed the total harvested energy, which incurs the EH constraint $(T-\tau_{i}^{(t)})p_{i}^{(t)}\leq E_{i}^{(t)}$.

Now, we jointly optimize the time allocation $\{\tau_{i}^{(t)}\}$ and the uplink power allocation $\{p_{i}^{(t)}\}$ to maximize the sum-rate performance. The problem is formulated as
\begin{align}\label{Problem}
    \max_{\substack{\{\tau_{i}^{(t)}\}},\{p_{i}^{(t)}\}}\frac{1}{T}\sum_{i=1}^{N}R_{i}^{(t)},
    \ \ s.t.\ (T-\tau_{i}^{(t)})p_{i}^{(t)}\leq E_{i}^{(t)}~~\forall i.
\end{align}
Problem \eqref{Problem} has been recently solved in \cite{HKim:18} based on traditional optimization techniques. Due to the combinatorial nature of the H-AP ordering, i.e., binary numbers in \eqref{crosslinkrate} and \eqref{Ein}, the sum rate maximization was carried out with an ordering $\tau_{1}\leq\cdots\leq\tau_{N}$ in \cite{HKim:18}. Therefore, the ordering should be additionally optimized in the outer loop, resulting in exhaustive search of size $N!$ whose computational complexity may become prohibitive for a large $N$. Furthermore, the conventional method in \cite{HKim:18} is based on the centralized computation where a central process unit is needed for the optimization of \eqref{Problem}. To address these issues, in this paper, by applying the RL technique, we propose a distributed optimization approach for \eqref{Problem} where H-AP $i$ individually obtains its solutions $\tau_{i}^{(t)}$ and $p_{i}^{(t)}$ while as well as the ordering in a single formulation.

\vspace{-1mm}
\section{Proposed MADRL Approach}\label{sec:proposed scheme}
This section presents the multi-agent advantage actor-critic (MA-A2C) method which handles \eqref{Problem} in a distributed manner. The A2C is a policy gradient based DRL technique which parameterizes a policy and updates the parameters of the policy to maximize the expected reward by the gradient method \cite{Sutton:14, Lei:19}. The A2C framework consists of an actor unit and a critic unit, which can be realized by DNN for a DRL setup. The actor determines a policy for an action of an agent, i.e., a solution of \eqref{Problem}, based on the states observed from the environment such as the channel gains. Utilizing the current policy, the critic estimates the expected reward value and helps the update of the actor. The A2C has been widely adopted for handling RL tasks with a large action set, possibly having infinitely many action candidates \cite{Sutton:14,Mnih:16}. Such a property is suitable for our formulation \eqref{Problem} which requires to find numerous combinations of the optimization variables $\tau_{i}^{(t)}$ and $p_{i}^{(t)}$, $\forall t,i$. Furthermore, in designing distributed RL strategies, a large action set is inevitable since it should allow agents to determine their own actions individually. This motivates us to apply the A2C approach to our WPCN formulation \eqref{Problem} which requires the optimization of $N^{2}$ dimensional space at each time slot. 

Now we explain how the A2C can estabilsh the distributed optimization of the WPCN. To this end, we present a multi-agent structure for the A2C framework as illustrated in Fig. \ref{figure:network}, where each agent is regarded as an H-AP responsible for a distributed decision of its resource allocation solution by using only locally observable information. Agent $i$ consists of an actor DNN and a critic DNN, each of which is represented by the parameters $\theta_{i}$ and $\phi_{i}$, respectively. The actor DNN of agent $i$ characterizes the conditional probability of the action $a_{i}^{(t)}$ for a given state $s_{i}^{(t)}$ denoted by the stochastic policy $\pi_{\theta_{i}}(a_{i}^{(t)}|s_{i}^{(t)})$ where the action $a_{i}^{(t)}$ is defined as a tuple $(\tau_{i}^{(t)},p_{i}^{(t)})$ of the time duration and the uplink power.

The critic DNN of agent $i$ models the value function $V_{\phi_{i}}^{\pi_{\theta_{i}}}(s_{i}^{(t)})$ under the policy $\pi_{\theta_{i}}$ which will be exploited for the optimization of the actor DNN \cite{Lei:19}. To be specific, we design the crtic DNN as a standard fully-connected DNN with a feedforward structure. On the other hand, for efficient learning of the joint stochastic policy $\pi_{\theta_{i}}(a_{i}^{(t)}|s_{i}^{(t)})$, we construct the actor DNN, where the input state is first pre-processed by several hidden layers, and then it is followed by two individual branches. Each of two branches determines the power and time allocation variables, respectively. In the following, we address the relationship between the problem in \eqref{Problem} and the MA-A2C structure by formulating the action, the state and the reward.

\vspace{-5mm}
\subsection{Action}\label{subsec:Action}
To construct a finite action set, we discretize the continuous variables $\tau_{i}^{(t)}$ and $p_{i}^{(t)}$. Specifically, for agent $i$, the action spaces $\mathcal{T}_{i}$ for the time allocation $\tau_{i}^{(t)}$ and $\mathcal{P}_{i}^{(t)}$ for the power allocation $p_{i}^{(t)}$ are respectively defined as
\vspace{-1mm}
\begin{align}
\mathcal{T}_{i}&=\ \bigg\{\frac{k(T-\epsilon_{\tau})}{K_{\mathcal{T}}-1} \ \text{for} \ k=0,1,\cdots,K_{\mathcal{T}}-1 \bigg\}\label{eq:Ti}\\
\mathcal{P}_{i}^{(t)}&=\bigg \{\frac{k E_{i}^{(t)}}{(K_{\mathcal{P}}-1)(T-\tau_{i}^{(t)})} \ \text{for} \ k=0,1,\cdots,K_{\mathcal{P}}-1 \bigg\}\nonumber
\end{align}
where a small positive number $\epsilon_{\tau}$ is introduced to avoid the case of $p_{i}^{(t)}\rightarrow\infty$, and $K_{\mathcal{T}}$ and $K_{\mathcal{P}}$ are the quantization level of time and power allocation, respectively. The size of the overall action space becomes $K_{\mathcal{T}}K_{\mathcal{P}}$.

The output of the actor DNN reflects the probability mass function of the elements in $\mathcal{T}_{i}$ and $\mathcal{P}_{i}^{(t)}$, and the action is then randomly sampled based on this probability mass function. It is worthwhile to note that a specific value of $E_{i}^{(t)}$ 
is required for the decision of $p_{i}^{(t)}$ at time slot $t$. This can be achieved as follows: First, all the H-APs transmit the WET signals in the downlink with the time duration $\tau_{i}^{(t)}$ generated from the actor DNN. Then, H-AP $i$ can measure $E_{i}^{(t)}$ from the received signal strength. With the estimated EH constraint $E_{i}^{(t)}$ at hands, the agents can determine the uplink power $p_{i}^{(t)}$ from the stochastic policy $\pi_{\theta_{i}}(a_{i}^{(t)}|s_{i}^{(t)})$.

\vspace{-3mm}
\subsection{State}\label{subsec:State}
The state $s_{i}^{(t)}$ of agent $i$ at time slot $t$ is constructed as a concatenation of the internal observation $s_{i,\text{I}}^{(t)}$ and the external information $s_{ji,\text{E}}^{(t)}$ sensed from all other cells $j\neq i$. Here, the internal information $s_{i,\text{I}}^{(t)}$ is defined as a collection of $a_{i}^{(t-1)}$, $H_{ii}^{(t-1)}$, $H_{ii}^{(t)}$, and $R_{i}^{(t-1)}$, which can be easily attained by an interaction between H-AP $i$ and user $i$. In contrast, $s_{ji,\text{E}}^{(t)}$ contains $\hat{E}_{ji}^{(t)}$, $\hat{I}_{ji}^{(t)}$, and $\hat{D}_{ji}^{(t)}$ defined as
\begin{align}
\hat{E}_{ji}^{(t)}&\!=\!
\eta P\!\!\sum_{n=1}^{N+1}(\tau_{\mu_{n}}^{(t\!-\!1)}\!-\!\tau_{\mu_{n-1}}^{(t\!-\!1)})h_{ij}^{(t)}(1\!-\!b_{ni}^{(t\!-\!1)})(1\!-\!b_{nj}^{(t\!-\!1)}),\label{eq:E_hat_ji}
\\\hat{I}_{ji}^{(t)}&=\sum_{n=1}^{N+1}(\tau_{\mu_{n}}^{(t-1)}-\tau_{\mu_{n-1}}^{(t-1)})h_{ji}^{(t)}p_{j}^{(t-1)}b_{nj}^{(t-1)},\label{eq:I_hat_ji}\\
\hat{D}_{ji}^{(t)}&=
\beta\sum_{n=1}^{N+1}(\tau_{\mu_{n}}^{(t-1)}-\tau_{\mu_{n-1}}^{(t-1)})g_{ji}^{(t)}(1-b_{nj}^{(t-1)}),\label{eq:D_hat_ji}
\end{align}
where $\hat{E}_{ji}^{(t)}$ stands for the estimation of the harvested energy $\sum_{n=1}^{N+1}E_{nji}^{(t)}$ in \eqref{Ein}, and $\hat{I}_{ji}^{(t)}$ and $\hat{D}_{ji}^{(t)}$ represent the WIT interference $\sum_{n=1}^{N+1}I_{nji}^{(t)}$ and the cross-link WET interference $\sum_{n=1}^{N+1}D_{nji}^{(t)}$ in \eqref{eq:interference} incurred by the $j$th interfering cell over all intervals, respectively. These quantities are measured over the current channel gains $h_{ji}^{(t)}$ and $g_{ji}^{(t)}$ following the past action $a_{j}^{(t-1)}$ of the $j$th interfering cell. As a consequent, agent $i$ can infer the actions of other agents $j$, $\forall j\neq i$, from its local state $s_{i}^{(t)}$ measured in a distributed manner.

The acquisition mechanism of the estimates \eqref{eq:E_hat_ji}-\eqref{eq:D_hat_ji} is given as follows: First, to get $\hat{E}_{ji}^{(t)}$ for $j\neq i$, H-AP $j$ transmits the WET signal at the beginning of time slot $t$ by using its previous action $a_{j}^{(t-1)}$, whereas other H-APs remain silent. Then, all the users readily obtain $\hat{E}_{ji}^{(t)}$ by observing the received signal power. Such a procedure is repeated $N$ times by changing the transmitting H-APs. Likewise, the interference levels $\hat{I}_{ji}^{(t)}$ and $\hat{D}_{ji}^{(t)}$ are obtained at H-AP $i$. Such a process needs no active data sharing among the agents, but depends on the sensing mechanism at the H-APs and users. Consequently, agent $i$ can build its state $s_{i}^{(t)}$ in a fully distributed manner.

\vspace{-3mm}
\subsection{Reward}
We present the reward $r_{i}^{(t+1)}$ of agent $i$ at time slot $t+1$ which should be maximized during the training process of the proposed MA-A2C scheme. Since our original target of \eqref{Problem} is to improve the overall data rate, the reward $r_{i}^{(t+1)}$ can be designed to include the current rate $R_{i}^{(t)}$. At the same time, we consider some penalizing terms since agent $i$ could degrade the network performance when it focuses only on its local rate $R_{i}^{(t)}$. To this end, we adopt the concept of a price for the power $p_{i}^{(t)}$ incurring interference to other cells \cite{Yasar:19}.

As a result, the reward $r_{i}^{(t+1)}$ is written by
\begin{align}\label{eq:reward}
    r_{i}^{(t+1)} = R_{i}^{(t)}-\sum_{j\neq i}^{N}(R_{j\backslash i}^{(t)}-R_{j}^{(t)}),
\end{align}
where the second term stands for the price preventing a naive decision $p_{i}^{(t)}=E_{i}^{(t)}/(T-\tau_{i}^{(t)})$. Here, $R_{j\backslash i}^{(t)}$ reflects the data rate of user $j$ achieved without interference from cell $i$ as 
\begin{align}
R_{j\backslash i}^{(t)}\!=\!
\sum_{n=1}^{N+1}(\!\tau_{\mu_{n}}^{(t)}\!-\!\tau_{\mu_{n\!-\!1}}^{(t)}\!)b_{nj}^{(t)}
\log\!\!\Bigg(\!\!1\!+\!\frac{h_{jj}^{(t)}p^{(t)}_{j}}{\sigma^{2}\!+\!\sum\limits_{k\neq i,j}(I_{nkj}^{(t)}\!+\!D_{nkj}^{(t)})}\!\Bigg).\nonumber
\end{align}

\vspace{-3mm}
\subsection{Learning and Implementation}\label{subsec:LearningStrategy}
We discuss a learning strategy for the proposed MA-A2C structure. The loss functions of the critic and actor DNNs of agent $i$ are respectively formulated as \cite{Sutton:14}
\begin{align}
    L_{i,C}&=\big((r_{i}^{(t+1)}+\gamma V_{\phi_{i}}^{\pi_{\theta_{i}}}(s_{i}^{(t+1)}))-V_{\phi_{i}}^{\pi_{\theta_{i}}}(s_{i}^{(t)})\big)^{2},\nonumber\\
    L_{i,A}&=-\log\pi_{\theta_{i}} (a_{i}^{(t)}|s_{i}^{(t)})\delta_{i}^{(t)},\nonumber
\end{align}
where $\gamma\in(0,1]$ denotes the discounting factor of the future rewards in the current time slot and $\delta_{i}^{(t)}\triangleq(r_{i}^{(t+1)}+\gamma V_{\phi_{i}}^{\pi_{\theta_{i}}}(s_{i}^{(t+1)}))-V_{\phi_{i}}^{\pi_{\theta_{i}}}(s_{i}^{(t)})$. Here, $L_{i,C}$ is set as the mean square error of the value function  $V_{\phi_{i}}^{\pi_{\theta_{i}}}(s_{i}^{(t)})$ and $L_{i,A}$ accounts for the cross-entropy of the stochastic policy $\pi_{\theta_{i}}(a_{i}^{(t)}|s_{i}^{(t)})$.

The critic and actor DNNs of agent $i$ are trained to minimize $L_{i,C}$ and $L_{i,A}$, respectively, based on the gradient descent (GD) algorithm. For  agent $i$, the update rules of the critic DNN $\phi_{i}$ and the actor DNN $\theta_{i}$ are written as \cite{Sutton:14}
\begin{align}
    \phi_{i}&\leftarrow \phi_{i}+\alpha_{C} \delta_{i}^{(t)}\triangledown_{\phi_{i}}V_{\phi_{i}}^{\pi_{\theta_{i}}}(s_{i}^{(t)}),\label{eq:phii}\\
    \theta_{i}&\leftarrow \theta_{i}+\alpha_{A}\delta_{i}^{(t)} \triangledown_{\theta_{i}}\log\pi_{\theta_{i}} (a_{i}^{(t)}|s_{i}^{(t)}),\label{eq:thetai}
\end{align}
where $\alpha_{C}$ and $\alpha_{A}$ are the learning rate for the critic and the actor, respectively. The updates in \eqref{eq:phii} and \eqref{eq:thetai} are carried out with a single state sample $s_{i}^{(t)}$, and thus no parallel computing capability is required.

We now explain the distributed training and execution processes of the proposed MA-A2C method. First, for the training, agent $i$ needs its current state $s_{i}^{(t)}$, future state $s_{i}^{(t+1)}$, and reward $r_{i}^{(t+1)}$ to proceed the GD updates \eqref{eq:phii} and \eqref{eq:thetai}. As mentioned in Sec. \ref{subsec:State}, the current state $s_{i}^{(t)}$ is readily obtained from the distributed sensing mechanism. Similarly, the future state $s_{i}^{(t+1)}$ can be collected at the beginning of time slot $t+1$. Then, the updates in \eqref{eq:phii} and \eqref{eq:thetai} are in fact performed right before the determination of the current actions. Finally, to compute the local reward in \eqref{eq:reward}, $R_{j}^{(t)}$ and $R_{j\backslash i}^{(t)}$ are locally calculated at H-AP $j$ and then forwarded to H-AP $i$ through backhaul connections. As a consequence, the distributed learning of the critic and actor DNNs is conducted by means of the locally sensed information as well as the knowledge obtained by other H-APs.

\begin{algorithm}[ht]
\caption{Distributed Training Algorithm}
\begin{algorithmic}
\State Initialize $s_{i}^{(t)}, \phi_{i}$ and $\theta_{i}, \ \forall i$ and set $t=0$
\NoDo
    \Repeat
        \State H-AP $i$, $\forall i$ takes $a_{i}^{(t)}$ using  its actor and $s_{i}^{(t)}$.
        \State H-AP $i$, $\forall i$ calculates $R_{i\backslash j}^{(t)}$ using $a_{i}^{(t)}$ and sends it
        \State \ to H-AP $j$, $\ \forall j\neq i$.
        \State H-AP $i$, $\forall i$ gets $r_{i}^{(t+1)}$ in (\ref{eq:reward}) and $s_{i}^{(t+1)}$ as in 
        \State \ Sec. \ref{subsec:State}.
        \State H-AP $i$, $\forall i$ updates $\phi_{i}$ and $\theta_{i}$ from (\ref{eq:phii}) and (\ref{eq:thetai})
        \State \  based on $r_{i}^{(t+1)}$ and $s_{i}^{(t+1)}$.
        \State $t\leftarrow t+1$
        \NoDo
    \Until{convergence}
\end{algorithmic}
\label{Algorithm 1}
\end{algorithm}

Algorithm \ref{Algorithm 1} summarizes the proposed distributed training procedure. The final goal of the algorithm is to implement individual updates in \eqref{eq:phii} and \eqref{eq:thetai}, which require $r_{i}^{(t+1)}$ and $s_{i}^{(t+1)}$ at H-AP $i$. First, to obtain the reward $r_{i}^{(t+1)}$, each H-AP takes the action $a_{i}^{(t)}$ using its actor and previous state $s_{i}^{(t)}$. From $a_{i}^{(t)}$, H-AP $i$ calculates $R_{i\backslash j}^{(t)}$ and sends it to other H-APs $j$, $\forall j\neq i$. In this way, H-AP $i$ can get $r_{i}^{(t+1)}$ by collecting the received information $R_{j\backslash i}^{(t)}$, $\forall j\neq i$. Next, to build $s_{i}^{(t+1)}$, H-AP $i$ broadcasts the WET signal using the previous action $a_{i}^{(t)}$. Then, user $j$, $\forall j\neq i$, measures the received signal power to obtain $\hat{E}_{ji}^{(t)}$. $\hat{I}_{ji}^{(t)}$ and $\hat{D}_{ji}^{(t)}$ are attained with a similar process as discussed in Sec. \ref{subsec:State}. Consequently, the reward $r_{i}^{(t+1)}$ and the state $s_{i}^{(t+1)}$ can be constructed in a distributed manner. This process is repeated during the training. 

The critic DNNs are discarded after the training. With the trained actor DNN at hand, the H-APs can decide the resource allocation strategy from the optimized stochastic policy function $\pi_{\theta_{i}} (a_{i}^{(t)}|s_{i}^{(t)})$. In the real-time execution step, it suffices for H-AP $i$ to know the local state $s_{i}^{(t)}$ for the individual decision. This leads to a distributed optimization of the multi-cell WPCN with arbitrary channel gains.

\vspace{-2mm}
\section{Simulation Results}
\label{sec:simulation}
We present numerical results to verify the proposed MA-A2C. The distance between an H-AP and its user and the distance among the H-APs are also set to $10$ m  and $15$ m, respectively \cite{HKim:18}, while the Rayleigh fading is assumed for small-scale channel gains. The path loss exponent is set to 3, and the maximum Doppler frequency and the duration of each time slot are respectively given by $f_{d}=10\ \text{Hz}$ and $T=20\ \text{ms}$ \cite{Yasar:19}. The H-APs utilize constant power budget $P=30\ \text{dBm}$. The WET signal cancellation factor and the noise power are fixed as $\beta=-50\ \text{dBm}$ and $\sigma^{2}=-50\ \text{dBm}$, respectively \cite{HKim:18}.

The critic DNN is constructed with four hidden layers each with 200, 200, 100 and 70 neurons. For the actor DNN, we consider two hidden layer with 200 neurons for the shared part, where two hidden layers with 200 neurons for each branch are subsequently connected. The hyperbolic tangent is applied as activations to all hidden layers.\footnote{We have numerically found that the hyperbolic tangent activations perform better than the rectifier linear unit or sigmoid activations.} Meanwhile, the linear and softmax functions are applied to output layers of the critic DNN and the actor DNN, respectively. The size of the action sets for the power and time variables are given as $K_{\mathcal{P}}=K_{\mathcal{T}}=20$, which corresponds to the size of the output layers for each branch of the actor DNN. The learning rate and the discount factor are fixed to $\alpha_{C}=\alpha_{A}=10^{-5}$ and $\gamma=0.5$, respectively. The training process lasts $10^{5}$ time slots, and it is followed by the testing of the trained DNNs with $10^{4}$ time slots. Both the training and testing performance are averaged over $50$ randomly initialized DNNs. Simulations are realized by Python and Tensorflow. 

\begin{figure}
\includegraphics[width=\FigureWidth]{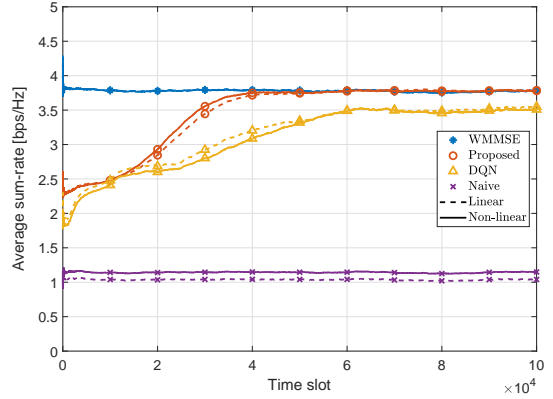}
\centering
\caption{Convergence of the distributed training process with $N=5$}
\label{figure:train performance}
\vspace{-2mm}
\end{figure}

Fig. \ref{figure:train performance} depicts the convergence behavior of the training process of the proposed MA-A2C method with $N=5$ cells for various EH models. The linear EH model is evaluated with $\eta=0.5$. The fitting parameters of the non-linear model in \eqref{eq:EH_NL} are set to $a_{1}=1.5\times 10^{3}$, $a_{2}=3.3$ and $a_{3}=2.8\times 10^{-3}$ \cite{Skang:19}. The following benchmarks are considered for the comparison.
\begin{itemize}
  \item \textit{Projected gradient decent (PGD)} \cite{HKim:18}: A PGD based alternating optimization algorithm is applied for determining locally optimum solutions $p_{i}^{(t)}$ and $\tau_{i}^{(t)}$ at each $t$ with the algorithm precision  $10^{-2}$.
  \item \textit{Baseline} \cite{Yasar:19}: The multi-agent deep Q-network architecture, which requires the centralized training process, is modified for our WPCN scenario. 
  \item \textit{Naive}: The users exhaust all the harvested energy with a simple equal time allocation strategy $\tau_{i}=T/2$, $\forall i$. 
\end{itemize}
We first observe that the average sum-rate performance of the proposed MA-A2C gradually converges to that of the locally optimal PGD algorithm. On the other hand, the baseline method, which relies on the centralized DNN implemented at each H-AP, fails to achieve the local optimal performance within $10^{5}$ time slots. This implies that the proposed MADRL architecture is crucial for achieving the optimal performance of the multi-cell WPCN. In addition, we can see that regardless of the EH models, the proposed MA-A2C shows a similar convergence behavior. In fact, the updates (\ref{eq:phii}) and (\ref{eq:thetai}) can be realized by observable state-action-reward tuples. Thus, the proposed method can adapt to arbitrary EH models.

\begin{figure}
      \centering
      \subfigure[Linear EH model]{
         \includegraphics[width=3.1in]{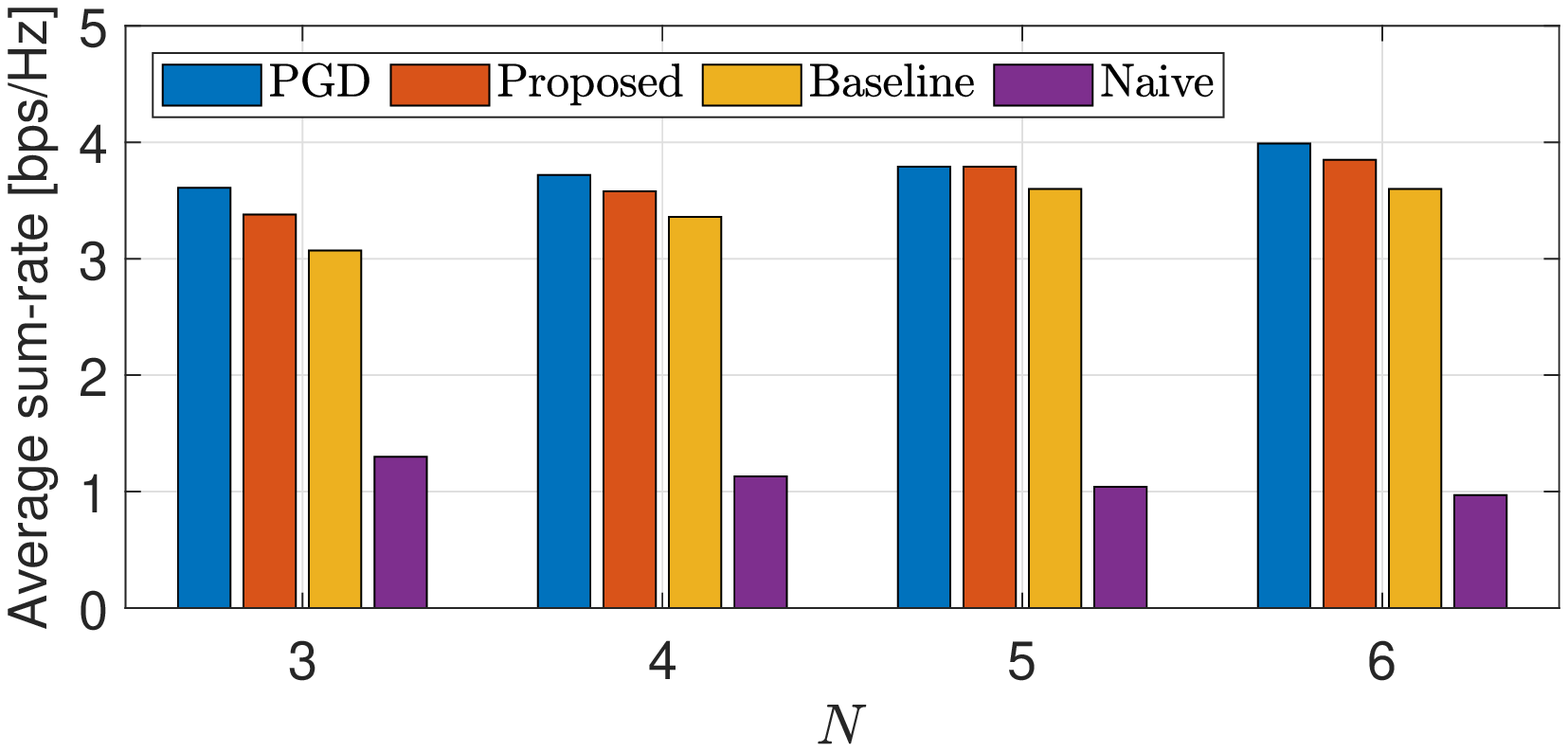}
         \label{figure:test performance L}}
      \subfigure[Non-linear EH model]{
         \includegraphics[width=3.1in]{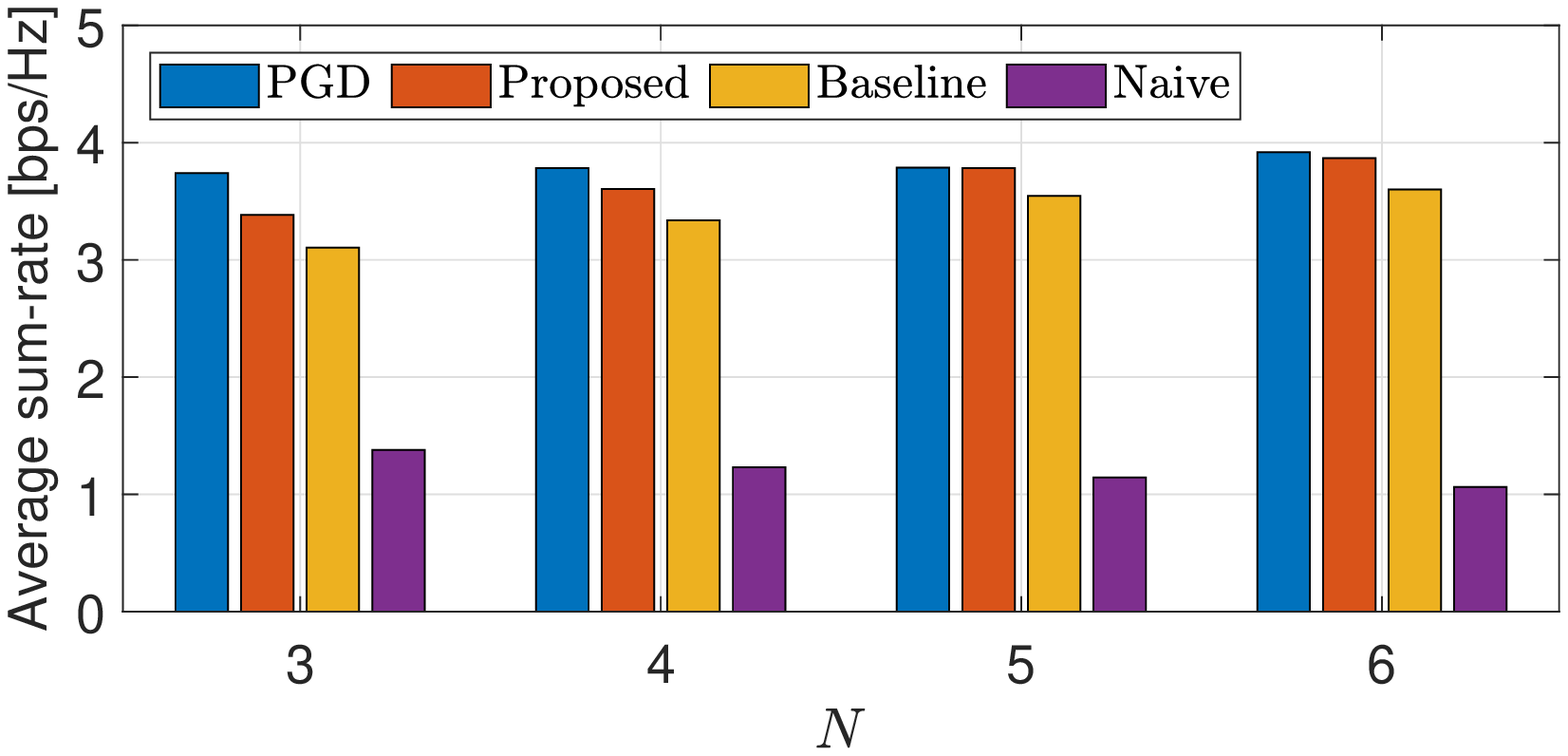}
         \label{figure:test performance NL}}
      \caption{Average sum-rate performance with respect to $N$}
      \label{figure:test performance}
      \vspace{-5mm}
   \end{figure}

Fig. \ref{figure:test performance} compares the average sum-rate with various $N$ for different EH models. Regardless of the EH models, the performance gap between the proposed MA-A2C and the PGD gets smaller as $N$ increases. It should be emphasized that the PGD is a centralized process, while our proposed scheme is based on a decentralized approach. Also, without the knowledge of specific EH models, our proposed scheme only exploits the measurable sensing information. The proposed MA-A2C becomes more efficient with a large $N$ by means of distributed coordinations among the agents. This verifies the effectiveness of the proposed local state and the distributed learning strategy. In addition, we can also see that the MA-A2C outperforms the baseline scheme over all $N$, verifying the effectiveness of the distributed MADRL structure. 

\begin{figure}
\includegraphics[width=\FigureWidth]{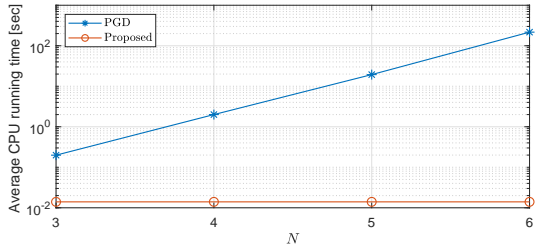}
\centering
\caption{Average CPU running time with respect to $N$}
\label{figure:CPU time}
\vspace{-5mm}
\end{figure}

Fig. \ref{figure:CPU time} exhibits the average CPU execution time of the proposed MA-A2C and the conventional PGD methods. The results are evaluated in Matlab R2019b on a PC equipped with an Intel Core i7-9700K @3.60 GHz processor with 16 GB RAM. Thanks to the distributed operations, H-APs optimized with the proposed MA-A2C can determine resource allocation variables in parallel, resulting in the same time complexity for all $N$. On the other hand, the PGD algorithm needs centralized optimization process coordinating, and thus its complexity rapidly increases as $N$ grows. This verifies the effectiveness of the distributed optimization structure for a large $N$.

\vspace{-3mm}
\section{Conclusion}
\label{sec:conclusions}
This paper has proposed a distributed optimization strategy for multi-cell WPCNs. A key idea is to develop a MA-A2C architecture so that each H-AP can determine its resource allocation solution in a distributed manner. To this end, we have carefully designed state variables at the H-APs by collecting locally observable statistics. Numerical results have verified the effectiveness of the proposed distributed optimization method. An extension to an energy-efficient design of multi-cell WPCN with the MA-A2C framework or a general multi-user setup would be an important future work.
\vspace{-2mm}

\bibliographystyle{ieeetr}
\begingroup
\renewcommand{\baselinestretch}{0.95}
\input{bibliography.filelist}
\endgroup

\end{document}